\documentclass[preprint,showpacs,showkeys,amsmath,amssymb,12pt,a4paper]{revtex4-1}
\usepackage[utf8]{inputenc}
\usepackage{graphicx}
\usepackage{amsmath}

\usepackage{color}
\usepackage{hyperref}
\begin{document}

\title{Perturbations to $\mu-\tau$ symmetry, lepton Number Violation and baryogenesis in left-right Symmetric Model}
\author{Happy Borgohain}
\email{happy@tezu.ernet.in}
\author{Mrinal Kumar Das}
\email{mkdas@tezu.ernet.in}
\affiliation{Department of Physics, Tezpur University, Tezpur 784028, India}

\begin{abstract}

 In this work, we studied baryogenesis via leptogenesis, neutrinoless double beta decay (NDBD) in the framework of LRSM where type I and type II
 seesaw terms arises naturally. The type I seesaw mass term is considered to be favouring $\mu-\tau$ symmetry, taking into account the widely
 studied  realizations of $\mu-\tau$ symmetric neutrino mass models, viz. Tribimaximal Mixing (TBM), Hexagonal Mixing (HM) 
 and Golden Ratio Mixing (GRM) respectively. The required correction to generate a non vanishing reactor mixing angle $\theta_{13}$ is obtained
 from the perturbation matrix, type II seesaw mass term in our case. We  studied the new physics contributions to NDBD and baryogenesis 
 ignoring the left-right gauge boson mixing and the heavy-light neutrino mixing, keeping mass of the gauge bosons and scalars to be around TeV 
 and studied the effects of the new physics contributions on the effective mass, NDBD half life and cosmological BAU and compared with the
 values imposed by experiments. We basically tried to find the leading order contributions to NDBD and BAU, coming from type
 I or type II seesaw in our work.
 \end{abstract}

\pacs{ 12.60-i, 14.60.Pq, 14.60.St}
\maketitle

\section{INTRODUCTION}

 The experimental evidence for neutrino oscillations coming from atmospheric, solar, reactor and long baseline neutrino experiments like
 MINOS \cite{minos}, T2K \cite{t2k}, Double Chooz \cite{doublechooz}, Daya Bay \cite{dayabay}, RENO \cite{reno} etc established the existence 
 of neutrino mass and large mixing parameters. In recent time, understanding the origin of Baryon Asymmetry of the universe (BAU) has been one of the most sought after  topic amongst the 
 scientific research community. It constitutes one of the major challenges in particle physics and cosmology and in our understanding of the dynamics
 of the universe. It appeals for a better explaination of the process beyond the most successful but inadequate Standard Model (SM) of particle physics.
 
 The quest for a better theory to complete the inconsistencies of SM gave rise to several models like the seesaw mechanism 
 (type I, type II, type III, Inverse seesaw) \cite{seesaw}, SUSY, extra dimensions, LRSM etc with some larger
 particle contents. Seesaw mechanism being the simplest way to understand the smallness of 
neutrino masses BSM. Nevertheless, LRSM is widely used and is an  appealing theory where the left and right chiralities are treated in equal footing
 at high energy scales. Herein, the Seesaw mechanisms arises naturally. The minimal LRSM is based on the gauge group 
 $ \rm SU(3)_c\times SU(2)_L\times SU(2)_R\times U(1)_{B-L}$, which extends the
 SM to restore parity as an exact symmetry which is then broken at some intermediate mass scale. The right handed (RH) neutrinos are a necessary 
 part of LRSM  which acquires a Majorana mass as soon as $SU(2)_R$ symmetry is broken at a scale $v_R$. Out of different mechanisms, leptogenesis
 is widely considered as favourable to explain BAU in the framework of LRSM. For leptogenesis to be testable in experiments, the breaking scale 
 of $SU(2)_R$ should be in the TeV range as well as there has to be a quasi degeneracy between at least two RH neutrinos with their mass difference 
 comparable to their decay widths for a resonant enhancement of the CP asymmetry. The connection between leptogenesis and low energy rare processes
 like Neutrinoless double beta decay (NDBD), LFV( Lepton Flavor Violation) etc cannot be overestimated. It has been extensively studied in 
 several earlier works \cite{connection}. Generally, the seesaw (SS) mechanism connects the light neutrinos with the heavy Majorana neutrinos, the decay of whose
 creates the leptonic CP asymmetry which can be converted to baryon asymmetry by the electroweak sphaleron transitions. In the scheme of LRSM, due to the presence of the heavy scalar particles, NDBD receives additional contributions 
 from RH gauge sector and the scalar triplets. Again, if the mass of the  scalar triplet is heavier than the lightest RH Neutrino mass which is of TeV scale,
 the asymmetry produced is dominated by the decay of only the RH neutrino and the leptogenesis from the decay of fermion or scalar triplet would also be ruled 
 out. In this case, leptogenesis can be explained from the type I SS diagrams, with
 the type I SS mass term, $ M_\nu\approx -M_DM^{-1}_{RR}M^T_D$, with a heavy-light neutrino mixing of order $M_D M^{-1}_{RR}$, where, $ M_D$ and $M_{RR}$ are
the Dirac and Majorana masses respectively. However, it would be enthralling to study the situation where both the situations (type I and type II
seesaw ) are comparable in size and the corresponding outcomes.
\par For a generic TeV scale LRSM, in order for neutrino mass to be of the order of sub eV, the Dirac Yukawa coupling has to be very small which would
lead to a very small efficiency factor in order for low scale leptogenesis to work. Several works have studied this in details and have come 
with some interesting outcomes that for maximal CP asymmetry (of order 1), successful BAU in LRSM requires the RH gauge boson mass to be greater than
18 TeV \cite{18tev}. Again, it has been found that for some specific textures of the Dirac and Majorana neutrino mass matrices, motivated by some flavor symmetry
in the leptonic sector, successful leptogenesis can also be realized for $M_{W_R}>$ 10 TeV with maximal CP asymmetry \cite{10tev}. The constraint on $W_R$ mass
is very important for the survival of LR SS leptogenesis. It would be interesting to probe the lower bound on $W_R$ mass to see if there exist any
allowed parameter space in TeV scale LRSS models with successful leptogenesis for smaller values of $M_{W_R}$.

\par In one of our earlier work, we have studied NDBD and BAU considering different  RH gauge boson mass 5, 10 and 18 TeV respectively and checked 
the consistency of several earlier results (second reference of \cite{connection}, considering equal contributions from both the type I and type II seesaw terms. In a recent paper, the 
authors have investigated if leptogenesis as a mechanism for explaining BAU can be tested at future colliders. They considered the case for two 
 RH Neutrinos in the mass range 5-50 GeV and estimated the allowed parameter space for successful leptogenesis \cite{50gev}.
However, the basic objective of this work is  to find  whether it is the type I or type II seesaw term which gives the leading contributions to NDBD and BAU. The RH gauge boson 
mass is considered to be 3.5 TeV, as accessible in experiments although we have shown in our previous work that  larger values of $M_{W_R}$ leads to
better efficiencies. With reference to one of our earlier work, we considered the type I SS mass term
to be favouring $\mu-\tau$ symmetry \cite{mutau}. The different realizations of $\mu-\tau$ symmetry which we have taken into account are Tribimaximal Mixing
(TBM), Hexagonal Mixing (HM) and Golden Ratio Mixing (GRM) \cite{ndbdlfv}. The perturbation to generate a non zero reactor mixing angle
is obtained from the type II seesaw term. It is quite natural to expect the model to have a lepton asymmetry once the $\mu-\tau$ symmetry
is broken, as well as also a non vanishing $\theta_{13}$.
The observation of NDBD would be significant as it would help us in understanding the origin of BAU as it would imply that lepton number indeed is not conserved
(one of the essential conditions for leptogenesis \cite{leptogenesis}). Furthermore, the Majorana nature \cite{maj} of neutrinos would also be established from NDBD. The latest experiments 
\cite{ndbd} that have improved the lower bound of the half life of the decay process include KamLAND-Zen \cite{kamland} and GERDA \cite{gerda} which uses Xenon-136 and 
Germanium-76 nuclei respectively. Incorporating the results from first and second phase of the experiment, KamLAND-Zen imposes the best lower limit on the decay half life using Xe-136 as
$ \rm T_{1/2}^{0\nu}>1.07\times 10^{26}$ yr at $ 90\%$ CL and the corresponding upper limit of effective Majorana mass in the range (0.061-0.165)eV.
\par In LRSM, there are several contributions to NDBD that involve left and right handed sectors individually as well as others that involve both sectors through left-right mixing
accompanied by both light and heavy neutrinos. Left-right mixing is always a ratio of the Dirac and Majorana mass scales $(M_DM^{-1}_{RR})$ which appears in the type I seesaw formula. NDBD involving left-right mixing can be 
enhanced for specific Dirac matrices. However in our present work, we have considered only two new physics contributions to NDBD coming from the diagrams containing purely RH 
 current mediated by the heavy gauge boson, $ \rm W_R$ by the exchange of heavy right handed neutrino, $ \rm N_R$ and another from the charged Higgs scalar $ \rm \Delta_R$ 
 mediated by the heavy gauge boson $ \rm W_R$ \cite{ndbddb2}. We have ignored the contributions coming from the left-right gauge boson mixing and heavy light 
 neutrino mixing owing to the very small left right mixing. 
\par This paper is organized as follows. In the next section, we present the left-right symmetric model framework with its particle contents and the origin of neutrino mass followed by the section \ref{sec:level4}, where we summarized
the implications of TeV scale LRSM in processes like BAU and other low energy observables like NDBD. In section \ref{sec:level5}, we present the basic steps
involved in our numerical analysis  and results and then give our conclusion  in section \ref{sec:level6}

 \section{LEFT RIGHT SYMMETRIC MODEL(LRSM) AND NEUTRINO MASS}{\label{sec:level3}}

 Left-right symmetric model (LRSM) has been extensively studied since 1970's by several groups (cite). Amongst the different extensions of the
 standard model (SM), the appealing and modest model which can be accessible at present day experiments
 is the LRSM \cite{LRSM} where the fermions are assigned to the gauge group  $ \rm SU(3)_c\times SU(2)_L\times SU(2)_R\times U(1)_{B-L}$ \cite{genericlrsm} \cite{LRSM} which
is a very simple extension of the standard model gauge group  where parity restoration is obtained at a high energy scale. The usual type I and II 
seesaw arises naturally in LRSM. The RH neutrinos are a necessary part of LRSM which acquires a Majorana mass when the 
$SU(2)_R$ symmetry is broken at a scale $v_R$. Several other problems 
like parity violation of weak interaction, masssless neutrinos, CP problems, hierarchy problems etc can also be addresses in the framework of LRSM.
The seesaw scale is identified as the breaking of the $SU(2)_R$ symmetry.
In this model, the electric charge generator is given by, $ Q=T_{3L}+T_{3R}+\frac{B-L}{2}$ \cite{Q}, where $ \rm T_{3L}$ and $ \rm T_{3R}$ are the generators of $ \rm SU(2)_L$ and $ \rm SU(2)_R$
and B-L being the baryon minus lepton number charge operator.

The Quarks and leptons (LH and RH) that transform in L-R symmetric gauge group are given by,

\begin{equation}\label{eqx}
Q^{'}_L=\left[\begin{array}{cc}
             u^{'}\\
            d^{'}
            \end{array}\right]_L,Q^{'}_R=\left[\begin{array}{cc}
             u^{'}\\
            d^{'}
            \end{array}\right]_R, \Psi^{'}_L=\left[\begin{array}{cc}
             \nu_l\\
            l
            \end{array}\right]_L,\Psi^{'}_R=\left[\begin{array}{cc}
             \nu_l\\
            l
            \end{array}\right]_R.
\end{equation}

where the Quarks are assigned with quantum numbers $(3,2,1,1/3)$ and $(3,1,2,1/3)$ and leptons with $(1,2,1,-1)$ and $(1,1,2,-1)$ respectively under
$ \rm SU(3)_c\times SU(2)_L\times SU(2)_R\times U(1)_{B-L}$. The Higgs sector in LRSM consists of a bi-doublet with quantum number $ \rm \phi(1,2,2,0)$
and the $SU(2)_{L,R}$ triplets, $\Delta_L(1,2,1,-1)$, $\Delta_R(1,1,2,-1)$. The matrix representation for which are given by,

\begin{equation}\label{eqx3}
\phi=\left[\begin{array}{cc}
             \phi_1^0 & \phi_1^+\\
             \phi_2^- & \phi_2^0
            \end{array}\right]\equiv \left( \phi_1,\widetilde{\phi_2}\right), \Delta_{L,R}=\left[\begin{array}{cc}
                     {\delta_\frac{L,R}{\sqrt{2}}}^+ & \delta_{L,R}^{++}\\
                     \delta_{L,R}^0 & -{\delta_\frac{L,R}{\sqrt{2}}}^+ .
                    \end{array}\right].
\end{equation}

The spontaneous symmetry breaking occurs in two successive steps given by,
$\rm SU(2)_L\times SU(2)_R\times U(1)_{B-L}\xrightarrow{<\Delta_R>} SU(2)_L\times U(1)_Y \xrightarrow{<\phi>} U(1)_{em}$. 
The VEVs  of the neutral component of the Higgs field are $ \rm v_R,v_L,k_1,k_2$ respectively. The VEV $ \rm v_R$ breaks the $ \rm SU(2)_R$ symmetry and sets the mass scale
for the extra gauge bosons $ \rm (W_R$ and Z$ \rm \ensuremath{'})$ and for right handed neutrino
field $ \rm (\nu_R)$. The VEVs $ \rm k_1$ and $ \rm k_2$ serves the twin purpose of breaking the remaining the  $ \rm SU(2)_L\times U(1)_{B-L}$ symmetry down to
$ \rm U(1)_{em}$, thereby setting 
the mass scales for the observed $ \rm W_L$ and Z bosons and providing Dirac masses for the quarks and leptons. Clearly, $ \rm v_R$ must be significantly larger 
than $ \rm k_1$ and $ \rm k_2$ in order for $ \rm W_R$ and Z $\ensuremath{'}$ to have greater masses than the $W_L$ and Z bosons. $v_L$ is the VEV of $\Delta_L$, 
it plays a significant role 
in the seesaw relation which is the characteristics of the LR model and can be written as,

\begin{equation}\label{eqx7}
 <\Delta_L>=v_L=\frac{\gamma k^2}{v_R}.
\end{equation}

\par The Yukawa lagrangian in the lepton sector is given by,
\begin{equation}\label{eqx8}
 \mathcal{L}=h_{ij}\overline{\Psi}_{L,i}\phi\Psi_{R,j}+\widetilde{h_{ij}}\overline{\Psi}_{L,i}\widetilde{\phi}\Psi_{R,j}+f_{L,ij}{\Psi_{L,i}}^TCi\sigma_2\Delta_L\Psi_{L,j}+f_{R,ij}{\Psi_{R,i}}^TCi\sigma_2\Delta_R\Psi_{R,j}+h.c.
\end{equation}

Where the family indices i,j are summed over, the indices $i,j=1,2,3$ represents the three generations of fermions. $C=i\gamma_2\gamma_0$ is the charge conjugation 
operator, $\widetilde{\phi}=\tau_2\phi^*\tau_2$ and $\gamma_{\mu}$ are the Dirac matrices. Considering discrete parity symmetry, the Majorana Yukawa couplings $f_L=f_R$ (for left-right symmetry) gives rises
to Majorana neutrino mass after electroweak symmetry breaking  when the triplet Higgs $\Delta_L$ and $\Delta_R$ acquires non zero vacuum expectation value.
Then equation (\ref{eqx8}) leads to $6\times6$ neutrino mass matrix as shown in reference 2 of \cite{ls1} 

\begin{equation}\label{eqx8}
M_\nu=\left[\begin{array}{cc}
              M_{LL}&M_D\\
              {M_D}^T&M_{RR}
             \end{array}\right],
\end{equation}

where 
\begin{equation}\label{eqx9}
M_D=\frac{1}{\sqrt{2}}(k_1h+k_2\widetilde{h}), M_{LL}=\sqrt{2}v_Lf_L, M_{RR}=\sqrt{2}v_Rf_R,
\end{equation}

where $M_D$, $M_{LL}$ and $M_{RR}$ are the Dirac neutrino mass matrix, left handed and right handed mass matrix respectively. Assuming $M_L\ll M_D\ll M_R$, the 
light neutrino mass, generated within a type I+II seesaw can be written as,

\begin{equation}\label{eqx10}
 M_\nu= {M_\nu}^{I}+{M_\nu}^{II},
\end{equation}

\begin{equation}\label{eqx11}
 M_\nu=M_{LL}+M_D{M_{RR}}^{-1}{M_D}^T
      =\sqrt{2}v_Lf_L+\frac{k^2}{\sqrt{2}v_R}h_D{f_R}^{-1}{h_D}^T,
\end{equation}

where the first and second terms in equation (\ref{eqx11}) corresponds to type II seesaw and type I seesaw mediated by RH neutrino respectively.
Here,
\begin{equation}\label{eqx12}
 h_D=\frac{(k_1h+k_2\widetilde{h})}{\sqrt{2}k} , k=\sqrt{\left|{k_1}\right|^2+\left|{k_2}\right|^2}.
\end{equation}

In the context of LRSM both type I and type II seesaw terms can be written in terms of $M_{RR}$ which arises naturally at a high energy scale as a result
of spontaneous parity breaking. In LRSM the Majorana Yukawa couplings $f_L$ and $f_R$ are same (i.e, $f_L=f_R$) and the VEV for left handed triplet $v_L$ can be written as,

\begin{equation}\label{eqx13}
v_L=\frac{\gamma {M_W}^2}{v_R}.
\end{equation}

Thus equation (\ref{eqx11}) can be written as ,

\begin{equation}\label{eqx14}
M_\nu=\gamma(\frac{M_W}{v_R})^2M_{RR}+M_D{M_{RR}}^{-1}{M_D}^T.
\end{equation}

In literature, (reference \cite{breaking} \cite{ndbddb2}) author define the dimensionless parameter $\gamma$ as,

\begin{equation}\label{eqx15}
 \gamma=\frac{\beta_1k_1k_2+\beta_2{k_1}^2+\beta_3{k_2}^2}{(2\rho_1-\rho_3)k^2}.
\end{equation}

Here the terms $\beta$, $\rho$ are the dimensionless parameters that appears in the expression of the Higgs potential.

\section{Resonant Leptogenesis and NDBD in TeV scale LRSM}{\label{sec:level4}}

Various models has been proposed and studied extensively for leptogenesis. For present day experiments accessible at LHCs, TeV scale SS models accounts for 
resonant leptogenesis \cite{RL}, a leptogenesis mechanism in which there is a resonant enhancement of the leptonic asymmetries when at least two heavy RH neutrinos 
are nearly degenerate and have mass difference comparable to their decay widths \cite{decaywidth}. In our analysis we considered the two nearly degenerate RH neutrinos to be of TeV 
range and have mass difference about $10^{-6}$ as per requirement of RL. In several earlier works, it hs been illustrated regarding the specific flavour 
structure that allows large Yukawa couplings that could serve the twin purpose of leptogenesis that could be efficient as well as testable in experiments.
Notwitstanding, as far as Dirac neutrino mass matrix is concerned we have not taken into account any specific structure of the matrix but  a generalized form
obtained by solving from the type I SS where the light neutrino mass matrix and the heavy RH Majorana mass matrix are considered to be known as in our previous 
work (second reference  of \cite{connection}). However, in this work, we have taken the type I mass term to be of different types obeying $\mu-\tau$ symmetry, notably, TBM, HM and GRM respectively
with reactor mixing angle, $\theta_{13}=0$. The perturbation to generate a non zero $\theta_{13}$ is obtained from the type II SS mass matrix, the elements of
which are explicitely shown in appendix section.

\par The underlying idea of this work is to relate leptonic asymmetry and hence baryon asymmetry  with low energy observable like NDBD as well as to find the 
leading order contribution on these phenomenon from the SS mechanisms (whether it is type I or type II SS). In LRSM, presence of the RH Neutrinos in type I SS and
the scalar triplets in type II SS propounds their decays that can give rise to lepton asymmetry. However, we would consider only the decay of the RH neutrinos  as in 
several earlier works and ignore the decay of the scalar triplets in generating leptonic asymmetry as above TeV scale , decay of the RH neutrinos are in thermal 
equillibrium and would wash out any premordial preceding leptonic asymmetries.
\par The heavy RH neutrinos present in the SS term besides explaining the origin of the tiny neutrino mass can also throw light on the cosmological baryonic 
asymmetry of the universe (obtained from the leptonic asymmetry and from EW sphaleron transistions). The lepton asymmetry is created by the decay of the heavy 
RH neutrinos into a lepton and a Higgs doublet $N_i\rightarrow L+\phi^{c} $ and its respective CP conjugate process,  $N_i\rightarrow L^c+\phi $
which can occur at both tree and one loop levels. The CP violating asymmetry $\epsilon_i$  arises from the interference between the tree level graph 
with absorptive part of self energy transition \cite{selfenergy} describing the mixing of the decaying particles and is defined as \cite{zing},

\begin{equation}\label{eqa7}
 \epsilon_i= \frac{\Gamma\left(N_i\rightarrow L+\phi^{c}\right)-\Gamma\left(N_i\rightarrow L^c+\phi\right)}{\Gamma\left(N_i\rightarrow L+\phi^{c}\right)+\Gamma\left(N_i\rightarrow L^c+\phi\right)}.
\end{equation}
which when satisfies the out of equillibrium condition can give rise to the required amount of lepton asymmetry.
 The decay rates of the  heavy neutrino decay processes are governed 
by the Yukawa couplings, and is given by,
\begin{equation}\label{eqa8}
 \Gamma_i= {\left(Y^\dagger_\nu Y_\nu\right)}_{ii}\frac{M_i}{8\pi}.
\end{equation}
 
For RL to occur, a prerequisite condition is $ M_i-M_j \approx \Gamma$ which leads to an enhancement of CP asymmetry even of order 1. Under such condition, 
RL can occur with heavy Majorana neutrinos even as light as $\approx $ 1 TeV. The  CP violating asymmetry $\epsilon_i$ is given by,

 \begin{equation}\label{eqa9}
 \epsilon_i= \frac{Im\left[{\left(Y^\dagger_\nu Y_\nu\right)}^2_{ij}\right]}{\left(Y^\dagger_\nu Y_\nu\right)_{11}\left(Y^\dagger_\nu Y_\nu\right)_{22}}.
 \frac{\left(M^2_i-M^2_j\right)M_i \Gamma_j}{\left(M^2_i-M^2_j\right)+M^2_i \Gamma^2_j},
\end{equation}

where,
\begin{equation}
 \frac{Im\left[{\left(Y^\dagger_\nu Y_\nu\right)}^2_{ij}\right]}{\left(Y^\dagger_\nu Y_\nu\right)_{11}\left(Y^\dagger_\nu Y_\nu\right)_{22}}\approx 1.
\end{equation}

The variables i, j run over 1 and 2, $i \neq j$.
\par The CP violating asymmetries $\epsilon_1$ and $\epsilon_2$ can give rise to a net lepton number asymmetry, provided the expansion rate of the universe is larger than
$\Gamma_1$ and $\Gamma_2$ at $T=M_N$. This can further be partially converted into baryon asymmetry of the universe by B+L violating sphaleron \cite{sphaleron} interactions which 
are in thermal equillibrium above the critical temperature $T_{c}$.

Presently one of the most preferred explanation of BAU emanates from lepton number violation or NDBD process which could prove the Majorana nature of the neutrinos.
NDBD plays a significant role to interpredt the dominance of matter over anti matter and the interrelation of NDBD nd BAU has been widely studid in many previous 
works. In the framework of LRSM there are various contributions to NDBD amplitude from the presence of several new heavy scalr particles apart from the standard
light neutrino contribution. It has been extensively studied in many of the earlier works (see ref. \cite{dbd1}\cite{ndbddb2}). The different new physics
contributions that could arise are coming from the ones mediated by $W_{R}$, the exchange of the heavy gauge bosons ( $ {W_L}^-$  and  $ {W_R}^-$ ),
both the left  and right handed gauge bosons (mixed diagrams, $\lambda$ and $\eta$) as well the scalar triplet ($\Delta_L$  and $\Delta_R$ ) contributions.
The amplitude of 
these processes mostly depends upon the mixing between light and heavy neutrinos, the  leptonic mixing matrix elements, the mass of the heavy neutrino ($M_i$), the mass of the gauge
bosons, ${W_L}^-$ and ${W_R}^-$ , the mass of the triplet Higgs  as well as their  coupling to leptons, $f_L$ and $f_R$ . 
\par However in our present work, we have considered only two of the aforesaid contributions to NDBD, the new physics contributions to NDBD that is the ones mediated by ${W_R}^-$ and $\Delta_R$ respectively.

\section{NUMERICAL ANALYSIS}{\label{sec:level5}}
Having studied several of the earlier works regarding NDBD and BAU in a TeV scale LRSM, in this work we are trying to do a comprehensive study of
these phenomenon within the framework of LRSM in the TeV scale, accessible in the colliders, encompassing the mostly studied $\mu-\tau$ symmetric 
neutrino mass models, namely, TBM, HM and GRM respectively. We took into consideration both the mass hierarchies, i.e, normal and inverted mass 
hierarchies. In our previous work, we studied BAU, NDBD and LFV and their correlation  by considering some specific values of right handed
gauge boson masses, 5, 10 and 18 TeV within and above the values measured in LHCs and checked the consistency of the results with several of the earlier
works.
\par Whereas, in this work we considered particularly the RH gauge boson mass to be 3.5 TeV as accessible in experiments and tried to study the contribution
of type I and type II seesaw with a view to finding the leading order and dominating contribution for BAU and NDBD. We have divided this section into two
subsections consisting of firstly the analysis of resonant leptogenesis and secondly of new physics contribution to NDBD.
\subsection{Baryogenesis via Leptogenesis}
Baryogenesis via leptogenesis has been widely studied in several earlier works. Herein, we are giving the detailed steps with relevant formulae involved
in the framework of LRSM in our analysis.
 
The light $\nu$ masses in the framework of LRSM can be written as,
\begin{equation}\label{eqa1}
 M_\nu={ M_\nu}^{I}+{ M_\nu}^{II},
\end{equation}
where,
\begin{equation}\label{eqa2}
{ M_\nu}^{I}= M_D{M_{RR}}^{-1}{M_D}^T.
\end{equation}
We have considered the type I mass term to be different realizations of $\mu-\tau$ symmetric neutrino mass models, tribimaximal mixing (TBM),
hexagonal mixing (HM) and golden ratio mixing (GRM) pattern (as in one of our previous work),
\begin{equation}\label{eqa3}
 { M_\nu}^{I}= U_{(\mu-\tau)}U_{Maj}{M_\nu}^{I(diag)}{U_{Maj}}^T{U_{(\mu-\tau)}}^T,
\end{equation}
where  $ \mu-\tau$ represents TBM, HM and GRM and

\begin{equation}
U_{TBM}=\left[\begin{array}{ccc}
             \frac{2}{\sqrt{6}} & \frac{1}{\sqrt{3}} & 0 \\
             -\frac{1}{\sqrt{6}} & \frac{1}{\sqrt{3}}& -\frac{1}{\sqrt{2}}\\
             -\frac{1}{\sqrt{6}} & \frac{1}{\sqrt{3}}& \frac{1}{\sqrt{2}}
            \end{array}\right], U_{HM}=\left[\begin{array}{ccc}
             \frac{\sqrt{3}}{2} & \frac{1}{2} & 0 \\
             -\frac{\sqrt{2}}{4} & \frac{\sqrt{6}}{4}& -\frac{1}{\sqrt{2}}\\
             -\frac{\sqrt{2}}{4} & \frac{\sqrt{6}}{4}& \frac{1}{\sqrt{2}}
            \end{array}\right], U_{GRM}=\left[\begin{array}{ccc}
             \frac{\sqrt{2}}{\sqrt{5-\sqrt{5}}} & \frac{\sqrt{2}}{\sqrt{5+\sqrt{5}}}& 0 \\
             -\frac{\sqrt{2}}{\sqrt{5+\sqrt{5}}} & \frac{\sqrt{2}}{\sqrt{5-\sqrt{5}}} & -\frac{1}{\sqrt{2}}\\
             -\frac{\sqrt{2}}{\sqrt{5+\sqrt{5}}} & \frac{\sqrt{2}}{\sqrt{5-\sqrt{5}}} &  \frac{1}{\sqrt{2}}
            \end{array}\right]
\end{equation}

$ \rm {M_\nu}^{I(diag)}=X{M_\nu}^{(diag)}$ \cite{X}, 
the parameter X describes the relative strength of the type I and II seesaw terms. It can take any numerical value
provided the two seesaw terms gives rise to correct light neutrino mass matrix. In our case, we considered three specific values of X, 
X = 0.3, 0.5 and 0.7 which corresponds to more contribution from type II, equal contribution from type I and type II seesaw and more contribution 
from type I seesaw respectively.\cite{X}
Thus, equation (\ref{eqa1}) can be written as,
\begin{equation}\label{eqa4}
 U_{PMNS}{M_\nu}^{(diag)} {U_{PMNS}}^T={ M_\nu}^{II}+U_{(\mu-\tau)}U_{Maj}X{M_\nu}^{(diag)}{U_{Maj}}^T{U_{(\mu-\tau)}}^T,
\end{equation}
$\rm U_{PMNS}$ being the diagonalizing matrix of the light neutrino mass matrix, $M_\nu$  given by,

\begin{equation}\label{eq5}
\rm U_{PMNS}=\left[\begin{array}{ccc}
c_{12}c_{13}&s_{12}c_{13}&s_{13}e^{-i\delta}\\
-c_{23}s_{12}-s_{23}s_{13}c_{12}e^{i\delta}&-c_{23}c_{12}-s_{23}s_{13}s_{12}e^{i\delta}&s_{23}c_{13}\\
 s_{23}s_{12}-c_{23}s_{13}c_{12}e^{i\delta}&-s_{23}c_{12}-c_{23}s_{13}s_{12}e^{i\delta}&c_{23}c_{13}
\end{array}\right]U_{Maj}.
\end{equation}

The abbreviations used are $c_{ij}$= $\cos\theta_{ij}$, $s_{ij}$=$\sin\theta_{ij}$, $\delta$ is the Dirac CP phase while the diagonal phase matrix,
$ \rm U_{Maj}$ is $ \rm diag (1,e^{i\alpha},e^{i(\beta+\delta)}) $, contains the Majorana phases $ \rm \alpha$ and $ \rm \beta$. The recent neutrino oscillation
data which we have adopted in our analysis is shown in the table \ref{t1},
\begin{table}[h!]
\centering
\begin{tabular}{||c| c| c||}
\hline
PARAMETERS & $3 \sigma$ RANGES & BEST FIT$\pm 1 \sigma$\\ \hline
$\Delta m_{21}^2[10^-5 \rm eV^2]$ & 7.05-8.14 & 7.56\\ \hline
$\Delta m_{31}^2[10^-3  \rm eV^2]$(NH) & 2.43-2.67 & 2.55\\
$\Delta m_{23}^2[10^-3 \rm eV^2]$(IH) & 2.37-2.61& 2.49\\ \hline
$\sin^2{\theta_{12}}$ & 0.273-0.379 & 0.321\\ \hline
$\sin^2{\theta_{23}}$(NH) & 0.384-0.635 & 0.430\\ 
(IH) & 0.388-0.638 & 0.596\\ \hline
$\sin^2{\theta_{13}}$(NH) & 0.0189-0.0239 & 0.0215\\ 
(IH) & 0.0189-0.0239 &  0.0214\\ \hline
$\delta/\pi$ & 0-2(NH)& 1.40\\
         & 0-2(IH)& 1.44\\ \hline
\end{tabular}
\caption{Global fit 3$\sigma$ values of $\nu$ oscillation parameters \cite{sigma}} \label{t1}
\end{table}

The RH Majorana neutrino mass $ \rm M_{RR}$ can be written in terms of type II SS mass term (from reference \cite{newphysics})as,
\begin{equation}\label{eqa5}
 M_{RR}=\frac{1}{\gamma}{\left(\frac{v_R}{M_{W_L}}\right)}^2{ M_\nu}^{II},
\end{equation}
Where, $\gamma$ is a dimensionless parameter and has been fine tuned as $\sim 10^{-10}$. As already mentioned we have considered  the $SU(2)_R$ 
breaking scale, $v_R$ to be specifically 3.5 TeV.  The left handed gauge boson mass is $M_{W_L}= 80$ GeV.

The type II SS mass term can be determined 
from the light neutrino mass and the type I SS mass term as,

\begin{equation}
 { M_\nu}^{II}= U_{PMNS}{M_\nu}^{(diag)} {U_{PMNS}}^T- U_{(\mu-\tau)}U_{Maj}X{M_\nu}^{(diag)}{U_{Maj}}^T{U_{(\mu-\tau)}}^T.
\end{equation}
The elements of the type II SS mass term is shown in appendix.
Again, $M_{RR}=U_R {M_{RR}}^{(diag)}{U_R}^T$,
${M_{RR}}^{(diag)}=diag(M_1, M_2, M_3)$. Expressing ${M_\nu}^{(diag)}$  in terms of lightest
neutrino mass, $m_1(m_3)$ for NH (IH), we obtained $M_{RR}$ varying the Majorana phases $\alpha$ and $\beta$ from 0 to 2$\pi$ and lightest neutrino mass
from $10^{-3}$ to $10^{-1}$. For leptogenesis to be testable in the colliders, i.e, for low scale 
leptogenesis, at least two of the lightest heavy RH neutrino has to have a very small mass difference and comparable to their decay widths. In such a case
the CP asymmetry can be resonantly enhanced. By considering a very tiny mass splitting of the Majorana masses
$M_1$ and $M_2$ as per requirement of resonant leptogenesis, we equated both sides of equation (\ref{eqa5}) and obtained  $M_1$, $M_2$ and  $M_3$,
where, $ M_1\approx M_2$.
\par As stated in literature (cite), a baryon asymmetry can be generated from a lepton asymmetry. We considered the lepton number violating and CP violating
out of equillibrium decays of two lightest heavy RH Majorana neutrinos, $N_1$ and  $N_2$ via the decay modes,  $N_i\rightarrow l+\phi^{c} $ and its 
CP conjugate process,  $N_i\rightarrow l^c+\phi $, $i=1, 2$. Firstly, we determined the leptonic CP asymmetry, $\epsilon_1$ and $\epsilon_2$
using equation (\ref{eqa9}) where $Y_\nu=\frac{M_D}{v}$, v being the VEV of Higgs bidoublet and is 174 GeV. The decay rates in equation 
(\ref{eqa9}) can be obtained using equation (\ref{eqa10}). 
The Dirac mass, $m_D$ generated due to neutrino Yukawa coupling, $Y_\nu$ after electroweak symmetry breaking can be determined from the type I
SS mass term provided the light neutrino mass and RH heavy neutrino mass is known. As mentioned before  $m_D$ is not of any specific texture, but we have obtained 
it from the type I SS equation  which satisfies the current neutrino oscillation data.

We have considered $M_D$ as,
\begin{equation}\label{b3}
\rm M_{D}=\left[\begin{array}{ccc}
a_1&a_2&a_3\\
a_2&a_4&a_5\\
a_3&a_5&a_6
\end{array}\right],
\end{equation}

which is $\mu-\tau$ symmetric. Equating both sides of type I seesaw equation and solving for $a_1,a_2,a_3,a_4,a_5,a_6$, we obtain 
the matrix elements of the $M_D$. One of $M_D$ for TBM, HM and GRM are obtained as,

\begin{equation}\label{b4}
\rm {M_{D}}^{TBM}=\left[\begin{array}{ccc}
-40246.3-40369.1i&49087.1-48889i&13531.9-18794.7i\\
49087.1-48889i&-59870-59391.8i&-16504.4-18996.2i\\
13531.9-18794.7i&-16504.4-18996.2i&-4549.79-85423.2i
\end{array}\right],
\end{equation}

\begin{equation}\label{b5}
\rm {M_{D}}^{HM}=\left[\begin{array}{ccc}
-57922+85072i&45976.8+55705.2i&-47286.5+39799.4i\\
45976.8+55705.2i&-36495.1+36544.7i&37534.7+24308.9i\\
-47286.5+39799.4i&37534.7+24308.9i&-38603.9+62978.5i
\end{array}\right],
\end{equation}

\begin{equation}\label{b5}
\rm {M_{D}}^{GRM}=\left[\begin{array}{ccc}
-46712.8-15424.2i&-58279+17674.6i&-12624.2-15616.3i\\
58279+17674.6i&-72709-55714.3i&-15749.9-29864i\\
-12624.2-15616.3i&-15749.9-29864i&-3411.69-80132i
\end{array}\right],
\end{equation}

which we have implemented for our further analysis.

 The CP violating asymmetries $\epsilon_1$ and $\epsilon_2$ can give rise to a net lepton number asymmetry, provided the expansion rate of the universe is larger than
$\Gamma_1$ and $\Gamma_2$. The net baryon asymmetry is then calculated using \cite{zing},

\begin{equation}\label{eqa10}
 \eta_B \approx -0.96 \times 10^{-2} \sum_i \left(k_i\epsilon_i\right),
\end{equation}

\begin{figure}[h!]
\includegraphics[width=0.25\textwidth]{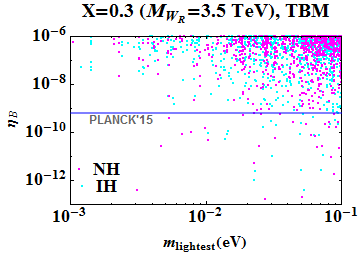}
\includegraphics[width=0.25\textwidth]{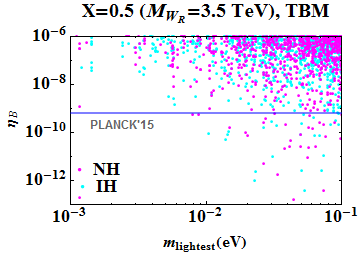}
\includegraphics[width=0.25\textwidth]{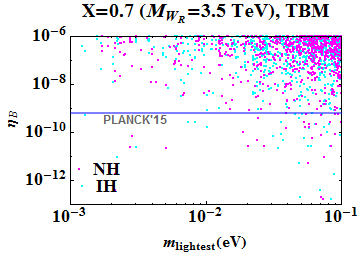}

\includegraphics[width=0.25\textwidth]{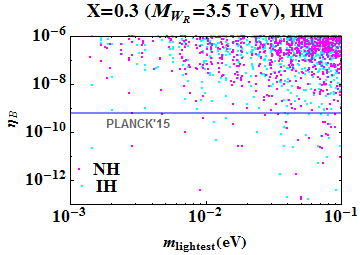}
\includegraphics[width=0.25\textwidth]{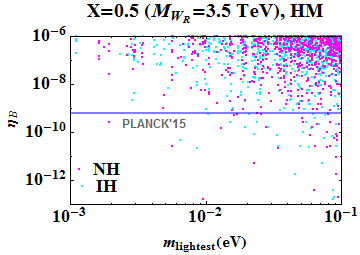}
\includegraphics[width=0.25\textwidth]{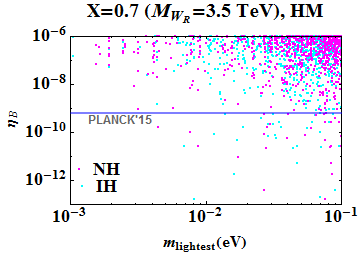}

\includegraphics[width=0.25\textwidth]{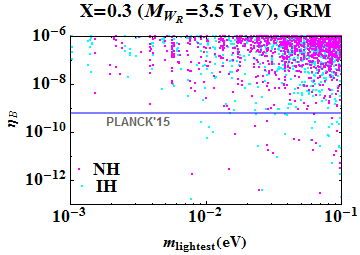}
\includegraphics[width=0.25\textwidth]{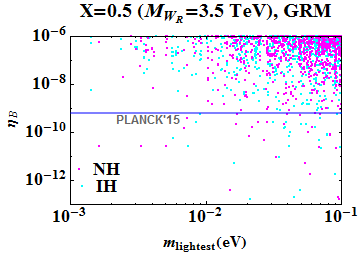}
\includegraphics[width=0.25\textwidth]{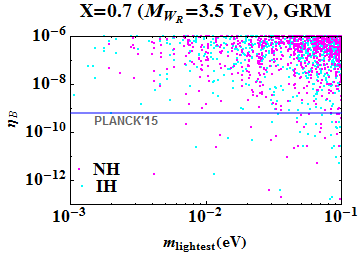}
\caption{BAU as a function of lightest neutrino mass, $m_1$ / $m_3$ for NH/IH for different neutrino mass models (TBM, HM, GRM) and different values of X.
The blue horizontal line indicates the bound on cosmological BAU from PLANCK '15} \label{fig4}
\end{figure}

\begin{figure}[h!]
\includegraphics[width=0.23\textwidth]{BAUTBMALPHA1.png}
\includegraphics[width=0.23\textwidth]{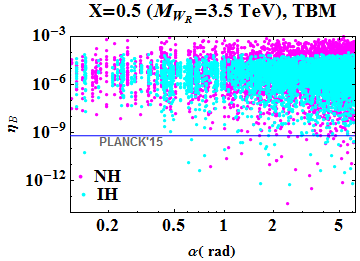}
\includegraphics[width=0.23\textwidth]{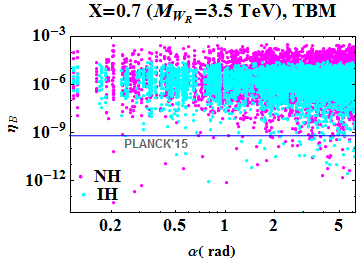}

\includegraphics[width=0.23\textwidth]{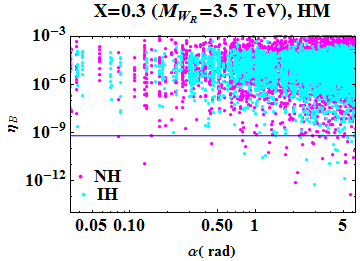}
\includegraphics[width=0.23\textwidth]{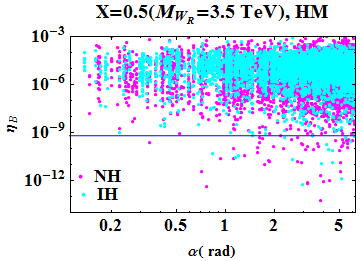}
\includegraphics[width=0.23\textwidth]{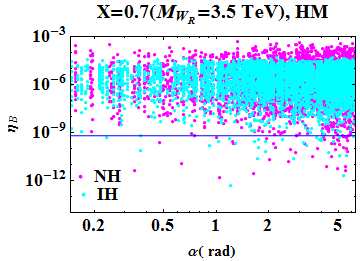}

\includegraphics[width=0.23\textwidth]{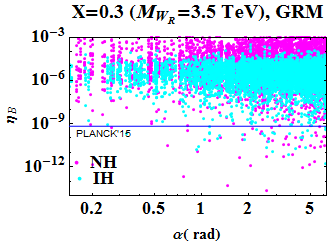}
\includegraphics[width=0.23\textwidth]{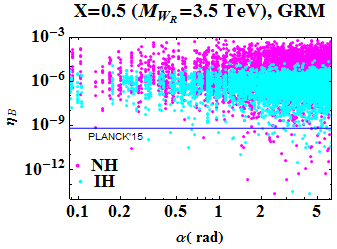}
\includegraphics[width=0.23\textwidth]{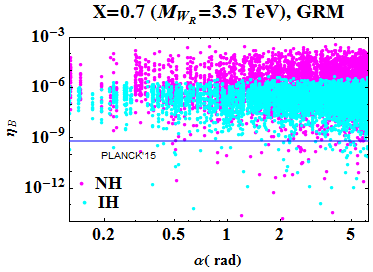}
\caption{BAU as a function of Majorana phase $\alpha$ for NH/IH for different neutrino mass models (TBM, HM, GRM). The blue horizontal line indicates the bound on cosmological BAU from PLANCK '15} \label{fig5}
\end{figure}

\begin{figure}[h!]
\includegraphics[width=0.23\textwidth]{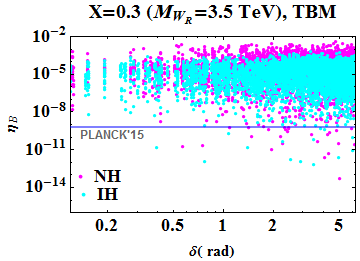}
\includegraphics[width=0.23\textwidth]{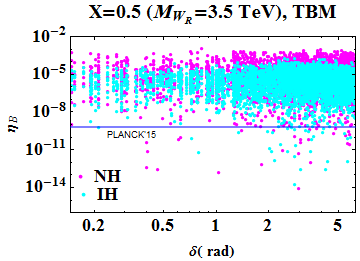}
\includegraphics[width=0.23\textwidth]{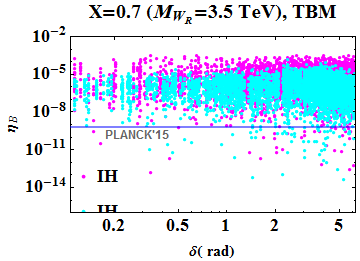}

\includegraphics[width=0.23\textwidth]{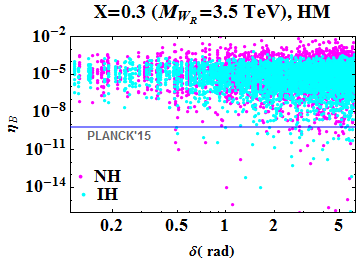}
\includegraphics[width=0.23\textwidth]{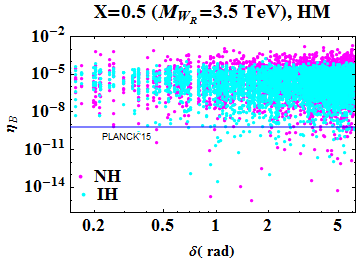}
\includegraphics[width=0.23\textwidth]{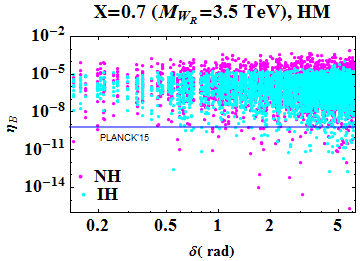}

\includegraphics[width=0.23\textwidth]{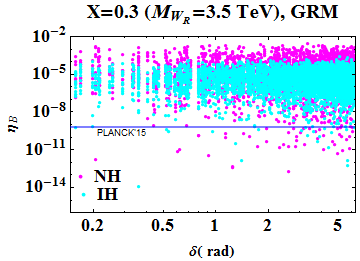}
\includegraphics[width=0.23\textwidth]{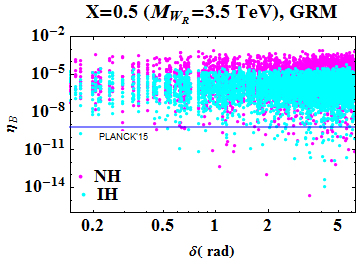}
\includegraphics[width=0.23\textwidth]{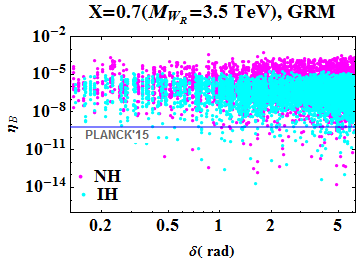}
\caption{BAU as a function of Dirac CP phase $\delta$ for NH/IH for different neutrino mass models (TBM, HM, GRM). The blue horizontal line indicates the bound on cosmological BAU from PLANCK '15} \label{fig6}
\end{figure}

$k_1$ and $k_2$ being the efficiency factors measuring the washout  effects linked with the out of equillibrium decay of $N_1$ and $N_2$. We can define the parameters,
$K_i\equiv \frac{\Gamma_i}{H}$ at temperature, $T= M_i$, $H\equiv \frac{1.66 \sqrt{g_*}T^2}{M_Planck}$ is the  Hubble's constant with $g_*\simeq$ 107 and 
$M_{Planck}\equiv 1.2 \times 10^{19} GeV$ is the Planck mass. The decay width can be estimated using equation (\ref {eqa8}). 
For simplicity, the efficiency factors, $k_i$ can be calculated using the formula \cite{K},
\begin{equation}\label{eqa11}
 k_1\equiv k_2\equiv \frac{1}{2}{\left(\sum_i K_i\right)}^{-1.2},
\end{equation}

 which holds validity for two nearly degenerate heavy Majorana  masses and $ 5\leq K_i\leq 100 $. We have used the formula \ref{eqa10} in calculating the baryon asymmetry.
 The result is shown as a function of lightest neutrino mass, Majorana phase $\alpha$ and Dirac phase $\delta$ in fig (\ref{fig4},\ref{fig5},\ref{fig6}) for different values of X.
It is evident from the figure that the cosmological observed BAU from RL can be obtained for different values of X, 0.3, 0.5 and 0.7.

\subsection{NDBD from heavy RH neutrino and scalar triplet contribution}

There are several new physics contributions to NDBD amplitudes due to the presence of the heavy scalar particles in LRSM. In the present work, we have considered
the contributions coming from the heavy RH neutrino contribution coming from the
exchange of $W_R$ bosons and from the scalar Higgs triplet. 

The effective neutrino mass corresponding to these conributions is given by,

\begin{equation}\label{eq55}
 \rm {m_{N+\Delta_R}}^{eff}=p^2\frac{{M_{W_L}}^4}{{M_{W_R}}^4}\frac{{U_{Rei}}^*2}{M_i}+p^2\frac{{M_{W_L}}^4}{{M_{W_R}}^4}\frac{{{U_{Rei}}^2}M_i}{{M_{\Delta_R}}^2}.
\end{equation}
\par Here, $ \rm <p^2> = m_e m_p \frac{M_N}{M_\nu}$ is the typical momentum exchange of the process, where $ \rm m_p$ and $ \rm m_e$ are the mass of the proton and electron respectively 
and $ \rm M_N$ is the NME corresponding to the RH neutrino exchange.
We know that TeV scale LRSM plays an important role in 0$\nu\beta\beta$ decay. We have considered the values $ \rm M_{W_R}$ = 3.5 TeV, $ \rm M_{W_L}$ = 80 GeV, $ \rm M_{\Delta_R}\approx $3TeV,
the heavy RH neutrino $\approx$ TeV which are within the recent collider limits. 
The allowed value of p, the virtuality of the exchanged neutrino is in the range
$\sim $ (100-200) MeV \cite{ndbddb2} and we have considered p$\simeq$180 MeV in our analysis as in earlier works.

\par Thus,
\begin{equation}\label{eq56}
 \rm p^2\frac{{M_{W_L}}^4}{{M_{W_R}}^4} \simeq 10^{10} {eV}^2.
\end{equation}
However, equation (\ref{eq55}) is valid only in the limit $ \rm {M_i}^2 \gg\left|<p^2>\right|$ and $ \rm {M_\Delta}^2\gg\left|<p^2>\right|$. 

To evaluate $ \rm {m_{N+\Delta_R}}^{eff}$, we need the
diagonalizing matrix of the heavy right handed Majorana mass matrix $ \rm M_{RR}$, $U_{Rei}$ and its mass eigenvalues, $ \rm M_i$.
 $ \rm M_{RR}$ can be written in the form (from reference \cite{newphysics}),
\begin{equation}\label{eq59}
 \rm M_{RR}=\frac{1}{\gamma}{\left(\frac{v_R}{M_{W_L}}\right)}^2{ M_\nu}^{II},
\end{equation}
\par For the new physics contribution in which the type II term acts as the perturbation, we have also evaluated
the half life of the $0\nu\beta\beta$ decay process using equation, 
\begin{equation}
\rm \Gamma^{0\nu}=\frac{1}{{T_{\frac{1}{2}}}^{0\nu}}=G^{0\nu}(Q,Z){\left|M^{0\nu}\right|}^2\frac{{\left|m_{\beta\beta}\right|}^2}{{m_e}^2}.
\end{equation}
 where
\begin{equation}\label{eq64}
\rm {\left| {m_\nu}^{eff}\right|}^2={\left|{m_N}^{eff}+{m_{\Delta_R}}^{eff}\right|}^2.
\end{equation}
Considering the values of the phase factors(${G_0}^\nu$) \cite{phasefactor} \cite{ndbdg}, nuclear matrix element(NME) \cite{nme} \cite{ndbdg} and mass
of electron, we have obtained the half life as a function of the lightest mass in the different mixing patterns for both NH and IH, as shown in figure (\ref{fig8}).

 We have summarized the plots of BAU and NDBD  for three different values of X as follows,

\begin{figure}[h!]
\includegraphics[width=0.23\textwidth]{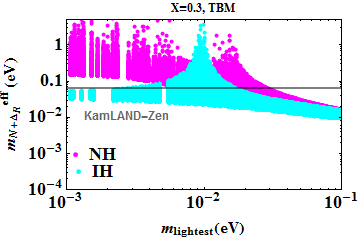}
\includegraphics[width=0.23\textwidth]{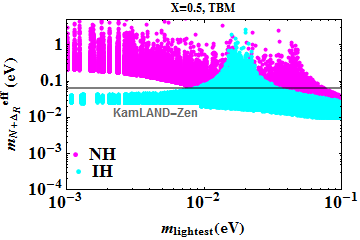}
\includegraphics[width=0.23\textwidth]{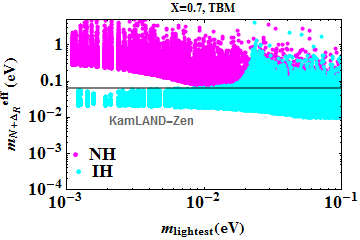}

\includegraphics[width=0.23\textwidth]{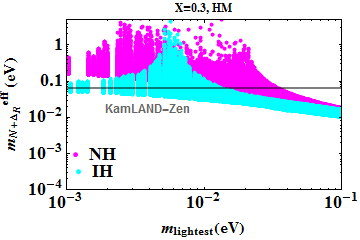}
\includegraphics[width=0.23\textwidth]{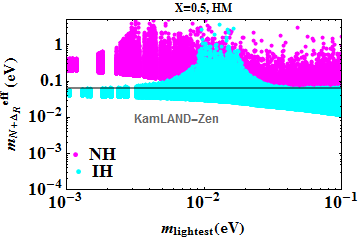}
\includegraphics[width=0.23\textwidth]{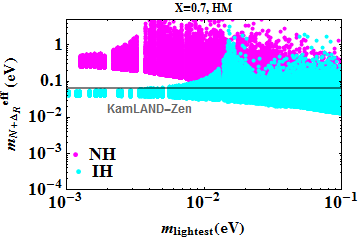}

\includegraphics[width=0.23\textwidth]{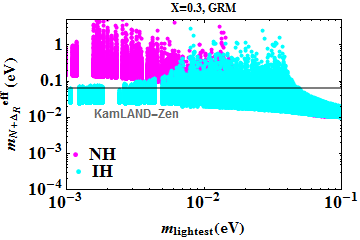}
\includegraphics[width=0.23\textwidth]{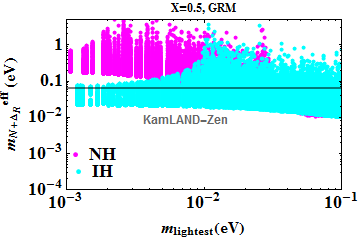}
\includegraphics[width=0.23\textwidth]{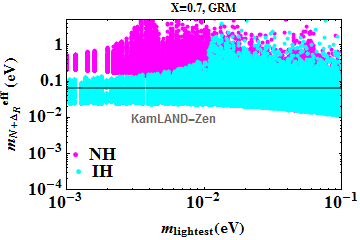}
\caption{New physics contribution to effective mass governing NDBD for different neutrino mass models for  NH and IH. The grey horizontal line indicates the upper limit on effective Majorana
mass given by KamLAND-Zen experiment.} \label{fig7}
\end{figure}

\begin{figure}[h!]
\includegraphics[width=0.23\textwidth]{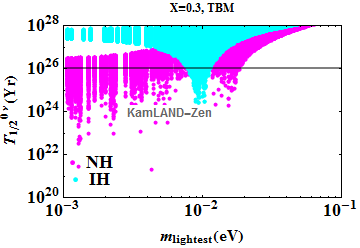}
\includegraphics[width=0.23\textwidth]{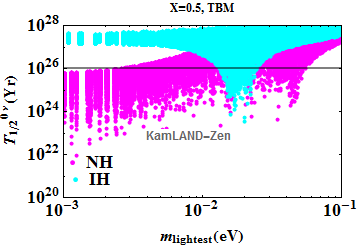}
\includegraphics[width=0.23\textwidth]{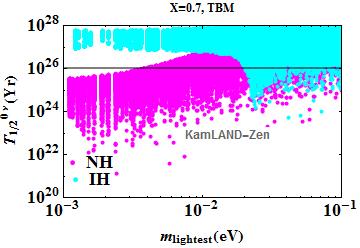}

\includegraphics[width=0.23\textwidth]{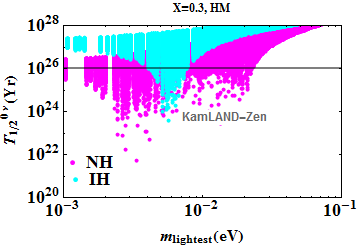}
\includegraphics[width=0.23\textwidth]{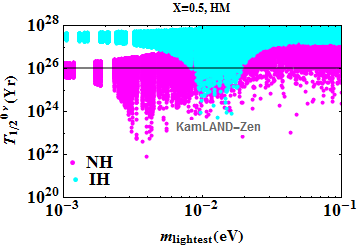}
\includegraphics[width=0.23\textwidth]{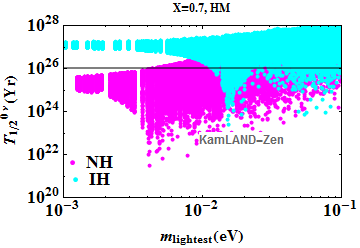}

\includegraphics[width=0.23\textwidth]{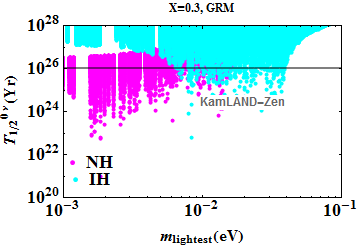}
\includegraphics[width=0.23\textwidth]{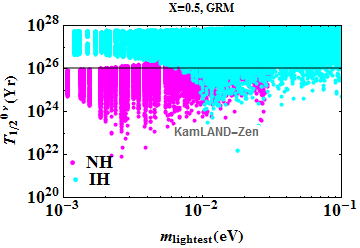}
\includegraphics[width=0.23\textwidth]{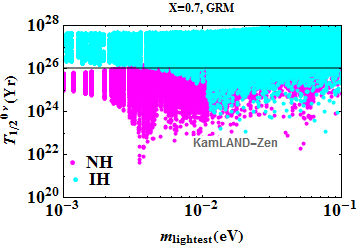}
\caption{New physics contribution to half life governing NDBD for different neutrino mass models for  NH and IH. The grey horizontal line indicates the lower limit on NDBD half life given by
KamLAND-Zen experiment.} \label{fig8}
\end{figure}

\clearpage

\section{DISCUSSION AND CONCLUSION}{\label{sec:level6}}
In this paper, we have done a phenomenological study of BAU and NDBD in the framework of TeV scale LRSM with the primary focus to see the contributions
of type I and type II SS terms to the abovementioned phenomenon considering both normal and inverted mass hierarchy of neutrino mass spectrum. In particular,
we have considered the type I SS mass term to be $\mu-\tau$ symmetric, namely, TBM, HM and GRM respectively whereas the perturbation  to generate non zero
$\theta_{13}$ has  been obtained from the type II SS term. It would be enthralling to explore the situations where both the contributions from type I and type II
SS are comparable in size or to speculate the dominance of either of the SS terms to study BSM phenomenon like, BAU and NDBD. Based on our study, we could arrive at the following conclusions,

\begin{itemize}

\item Successful leptogenesis can be accounted for, considering $M_{W_R}$ as low as 3.5 TeV for a model independent analysis irrespective of different mixing 
patterns.

 \item The baryon asymmetry, $\eta_B$ is found in the observable range in all the cases irrespective of the mass hierarchies and the type 
 I/II seesaw contribution. Most of the observed values are found in the lightest mass range varying from $10^{-2}$ to $10^{-1}$ eV.

 \item $\eta_B$ Vs $\alpha$ plot shows us that IH tends to be close to the observed value of $\eta_B$  although both the hierarchies are consistent
 with the experimentally observed value. We cannot conclude much about the leading contribution from the figure.
 
  \item The variation of $\eta_B$ with the Dirac CP phase $\delta$ shows equal dominance of both the mass hierarchies irrespective of the neutrino mass models,
 TBM, HM and GRM.
 
  \item In new Physics contributions to NDBD in TeV scale LRSM, TBM, HM, GRM shows results within experimental bound for a wide range of lightest
 neutrino mass varying from $10^{-3}$ to $10^{-1}$, however for X= 0.7 (i.e leading type II contribution), the results are widely 
 scattered for lightest neutrino mass varying from $10^{-2}$ to $10^{-1}$. For NH, effective mass lies within bound for $m_{lightest}$ varying from 
 ($10^{-2}$ to $10^{-1}$) eV.
 \par Similarly, for half life, IH gives a better result in all the cases (TBM, HM, GRM) as far as experimental bounds are concerned.

\end{itemize}

\section{APPENDIX}{\label{sec:level7}}

\textbf{Elements of the type II Seesaw mass matrix:}

\begin{equation}
 S_{11}=\left(c^2_{12}c^2_{13}-X{c_{12}^2}^{\mu\tau}\right)m_1+e^{2i(\beta-\delta)}s^2_{13}m_3+\left(c^2_{13}s^2_{12}-X{s_{12}^2}^{\mu\tau}\right)e^{2i\alpha}m_2
\end{equation}
\begin{equation}
\begin{split}
S_{12}=\left(-c_{12}c_{13}c_{23}s_{12}-c^2_{12}c_{13}s_{13}s_{23}e^{i\delta}+X{c_{12}^{\mu\tau}}{c_{23}^{\mu\tau}}{s_{12}^{\mu\tau}}\right)m_1+\\
\left(-c_{13}s_{12}c_{12}c_{23}e^{2i\alpha}-c_{13}s^2_{12}s_{13}s_{23}e^{i(2\alpha+\delta)}+X{c_{12}^{\mu\tau}}{c_{23}^{\mu\tau}}{s_{12}^{\mu\tau}}e^{2i\alpha}\right)m_2+\\
\left(c_{13}s_{13}s_{23}e^{i(2\beta-\delta)}\right)m_3
\end{split}
\end{equation}
\begin{equation}
\begin{split}
 S_{13}=\left(c^2_{12}c_{13}c_{23}s_{13}e^{i\delta}+s_{12}s_{23}c_{12}c_{13}-X{c_{12}^{\mu\tau}}{s_{12}^{\mu\tau}}{s_{23}^{\mu\tau}}\right)m_1+\\
 \left(-c_{13}s_{12}c_{23}s_{12}s_{13}e^{i(2\alpha+\delta)}-X{c_{12}^{\mu\tau}}{s_{12}^{\mu\tau}}{s_{23}^{\mu\tau}}e^{2i\alpha}\right)m_2+\\
 \left(e^{i(2\beta-\delta)}c_{13}c_{23}s_{13}\right)m_3
 \end{split}
\end{equation}
\begin{equation}
\begin{split}
 S_{21}=\left(-c_{12}c_{13}c_{23}s_{12}-c^2_{12}c_{13}s_{13}s_{23}e^{i\delta}+X{c_{12}^{\mu\tau}}{c_{23}^{\mu\tau}}{s_{12}^{\mu\tau}}\right)m_1+\\
 \left(c_{13}s_{12}c_{12}c_{23}e^{2i\alpha}-s^2_{12}s_{13}s_{23}c_{13}e^{i(2\alpha+\delta)}+X{c_{12}^{\mu\tau}}{c_{23}^{\mu\tau}}{s_{12}^{\mu\tau}}e^{2i\alpha}\right)m_2\\
 \left(e^{i(2\beta-\delta)}c_{13}s_{23}s_{13}\right)m_3
 \end{split}
\end{equation}
\begin{equation}
 \begin{split}
  S_{22}=\left({\left(c_{23}s_{12}-e^{i\delta}c_{12}s_{13}s_{23}\right)}^2-X{c_{23}^2}^{\mu\tau}{s_{12}^2}^{\mu\tau}\right)m_1+\\
  \left(-X{c_{12}^2}^{\mu\tau}{c_{23}^2}^{\mu\tau}+{\left(-c_{12}c_{23}-e^{i\delta}s_{12}s_{13}s_{23}\right)}^2\right)m_2e^{2i\alpha}+\\
  \left(c^2_{13}s^2_{23}-X{s_{23}^2}^{\mu\tau}e^{2i\beta}\right)m_3
  \end{split}
\end{equation}
\begin{equation}
 \begin{split}
S_{23}= \left(\left(-c_{12}c_{23}s_{13}e^{i\delta}+s_{12}s_{23}\right)\left(-c_{23}s_{12}-e^{i\delta}c_{12}s_{13}s_{23}\right)+X{c_{23}^{\mu\tau}}{s_{12}^2}^{\mu\tau}{s_{23}^2}^{\mu\tau}\right)m_1+\\
 \left(\left(-e^{i\delta}c_{23}s_{12}s_{13}+c_{12}s_{23}\right)\left(-c_{12}c_{23}-e^{i\delta}s_{12}s_{13}s_{23}\right)+X{c_{12}^2}^{\mu\tau}{c_{23}^{\mu\tau}}{s_{23}^{\mu\tau}}\right)m_2e^{2i\alpha}+\\
 \left(c^2_{13}c_{23}s_{23}e^{2i\beta}-{c_{23}^{\mu\tau}}{s_{23}^{\mu\tau}}\right)m_3
 \end{split}
\end{equation}
\begin{equation}
 \begin{split}
S_{31}=\left(c^2_{12}c_{13}c_{23}s_{13}e^{i\delta}+s_{12}s_{23}c_{12}c_{13}-X{c_{12}^{\mu\tau}}{s_{12}^{\mu\tau}}{s_{23}^{\mu\tau}}\right)m_1+\\
  \left(c_{13}s^2_{12}e^{i\delta}c_{23}s_{13}+c_{12}s_{23}c_{13}s_{12}e^{2i\alpha}-X{c_{12}^{\mu\tau}}{s_{12}^{\mu\tau}}{s_{23}^{\mu\tau}}\right)m_2e^{2i\alpha}+\\
  \left(e^{2i\beta-i\delta}c_{13}c_{23}s_{13}\right)m_3
 \end{split}
\end{equation}
\begin{equation}
 \begin{split}
S_{32}=\left(\left(-e^{i\delta}c_{12}c_{23}s_{13}+s_{12}s_{23}\right)\left(-c_{23}s_{12}-e^{i\delta}c_{12}s_{13}s_{23}\right)+{c_{23}^{\mu\tau}}{s_{12}^2}^{\mu\tau}{s_{23}^{\mu\tau}}\right)m_1\\
 \left(\left(-e^{i\delta}c_{23}s_{12}s_{13}+c_{12}s_{23}\right)\left(-c_{12}c_{23}-e^{i\delta}s_{12}s_{13}s_{23}\right)+X{c_{12}^2}^{\mu\tau}{c_{23}^{\mu\tau}}{s_{23}^{\mu\tau}}\right)e^{2i\alpha}m_2\\
  \left(c^2_{13}c_{23}s_{23}-X{c^{\mu\tau}}_{23}{s^{\mu\tau}}_{23}\right)e^{2i\beta}m_3
 \end{split} 
\end{equation}
\begin{equation}
 \begin{split}
S_{33}=\left({\left(-e^{i\delta}c_{12}c_{23}s_{13}+s_{12}s_{23}\right)}^2-X{s_{12}^2}^{\mu\tau}{s_{23}^2}^{\mu\tau}\right)m_1+\\
  \left({\left(-e^{i\delta}c_{23}s_{12}s_{13}+c_{12}s_{23}\right)}^2-X{c_{12}^2}^{\mu\tau}{s_{23}^2}^{\mu\tau}\right)e^{2i\alpha}m_2+\\
  \left(c^2_{13}c^2_{23}-{c_{23}^2}^{\mu\tau}\right)e^{2i\beta}m_3
 \end{split}
\end{equation}
Where, $\rm {c_{ij}^{\mu\tau}}= \cos\theta_{ij}^{\mu\tau}$, $\rm {s_{ij}^{\mu\tau}}=\sin\theta_{ij}^{\mu\tau}$ represents the mixing angles for $\rm \mu-\tau$ symmetric neutrino mass matrix (TBM, HM, GRM).

\end{document}